%% file: DER_allocation_for_storm-induced_failures_in_DNs.tex
\documentclass[conference]{IEEEtran}

\usepackage[letterpaper,top=0.75in,left=0.74in,right=0.74in,bottom=0.75in]{geometry}

\input{sources/preamble}

\input{sources/macros}

\graphicspath{ {./figuresSGC_Rev/} }

\begin{document}

\title{DER Allocation and Line Repair Scheduling for Storm-induced Failures in Distribution Networks}

\IEEEoverridecommandlockouts
	\author{Derek Chang, Devendra Shelar, and Saurabh Amin
	\thanks{Mailing address: Massachusetts Institute of Technology, 77 Massachusetts Avenue 1-241, Cambridge, MA 02139 USA (e-mail: \texttt{changd,shelard,amins}@mit.edu, phone: 857-253-8964).}
	\thanks{This work was supported by NSF CAREER award CNS 1453126, NSF FORCES award CNS-1239054, and the National Science Foundation Graduate Research Fellowship under Grant No. 1122374.}
}
\lhead{Optimal Reserve Allocation}


\maketitle

\begin{abstract}
	Electricity distribution networks (DNs) in many regions are increasingly subjected to disruptions caused by tropical storms.  Distributed Energy Resources (DERs) can act as temporary supply sources to sustain \enquote{microgrids} resulting from disruptions. In this paper, we investigate the problem of suitable DER allocation to facilitate more efficient repair operations and faster recovery. First, we estimate the failure probabilities of DN components (lines) using a stochastic model of line failures which parametrically depends on the location-specific storm wind field. Next, we formulate a two-stage stochastic mixed integer program, which models the distribution utility's decision to allocate DERs in the DN (pre-storm stage); and accounts for multi-period decisions on optimal dispatch and line repair scheduling (post-storm stage).  A key feature of this formulation is that it jointly optimizes electricity dispatch within the individual microgrids and the line repair schedules to minimize the sum of the cost of DER allocation and cost due to lost load. To illustrate our approach, we use the sample average approximation method to solve our problem for a small-size DN under different storm intensities and DER/crew constraints.    
\end{abstract}


%

\section{Introduction}\label{sec:introduction}
Weather-related outages in electricity distribution networks (DNs) continue to show an upward trend as utilities face the dual problems of deteriorating power grid infrastructure and higher frequency of natural disasters such as hurricanes~\cite{Campbell,Davidson1}.
Prolonged delays in restoring the power supply for Puerto Rico in the aftermath of Hurricane Maria highlight the importance of strategic planning and efficient response to extreme events. This paper is motivated by the need for developing a modeling framework that (i) accounts for the likely locations of component failures for damage assessment; and (ii) enables the design of pre-storm resource allocation strategies as well as post-storm repair operations. To address these issues, we formulate a two-stage stochastic optimization problem based on an uncertainty model of storm-induced failures. 

Our uncertainty model utilizes predictions of storm tracks and surface wind velocities over a spatial region during the expected duration of the storm (see \cite{BaoFuentes} for a related approach). Hours or days in advance of a storm, one can obtain track forecasts from public sources such as the National Hurricane Center (NHC). For a forecasted storm track, we estimate the surface wind velocity field using well-known parametric models \cite{Holland}. We focus on wind-induced damage (as opposed to flooding-induced failures), as strong winds during a storm are reported to be one of the primary factors for failures of above ground DN components, such as the failures resulting from falling of trees/vegetation on power lines and poles~\cite{LiGengfeng}. Next, we estimate the failure probabilities of DN components using a non-homogeneous Poisson process (NHPP) model, which depends on the estimated location-specific wind velocities \cite{Zhou}-\cite{AS}. Finally, these failure probabilities are used as an input to the two-stage stochastic optimization problem.

A key aspect of our formulation is that we allow for the partial DN operation in situations when the bulk power supply is no longer available, and it is beneficial to operationalize microgrids during the recovery period. Literature is now available on the allocation of repair crew and optimal response operations \cite{vanHentenryck, Whipple, Perakis, Kirschen}. These contributions focus on resource limitations, failure uncertainties, and physical constraints. However, the problem of proactive allocation of temporary generators in the pre-storm stage has received limited attention in the literature. This opportunity becomes especially relevant given the technological progress in portable Distributed Energy Resources (DERs) and  microgrid technologies~\cite{hurricaneMicrogrid}. The significance of proactive DER allocation in the face of natural disasters has already been acknowledged by federal agencies~\cite{derSiteFEMAArticle,derSiteArticle}. Our formulation allows for strategic placement of DERs at a subset of DN nodes in the pre-storm stage, given the uncertainty in component failures and the resulting outages and lost load for a particular storm. These DERs can then be used to sustain microgrids in the post-storm stage~\cite{Badolato}, while the line repair operations are being completed and the connection to bulk supply is being restored. 

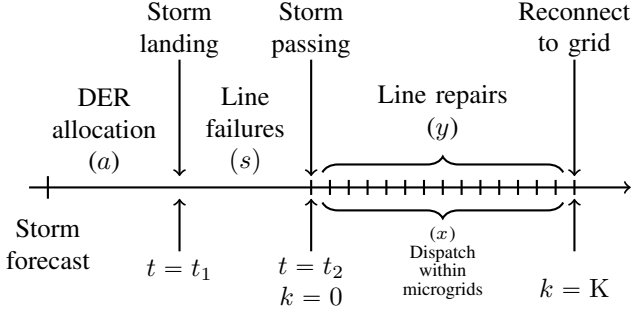
\begin{figure}[htbp!]
	\centering
	\begin{tikzpicture}[scale=1]
	\draw [thick,-] (0.25,0.15) -- (.25,-0.15);
	\node[align=center] at (1,0.75) {DER\\allocation\\($\first$)};
	\node[align=center] at (.25,-0.75) {Storm\\forecast};
	\node[align=center] at (2,2.1) {Storm\\landing};
	\node[align=center] at (2,-1.1) {$\time=\time_1$};
	\draw [thick,->] (2,-0.85) -- (2,-0.15);
	\draw [thick,->] (2,1.7) -- (2,0.15);
	\node[align=center] at (2.875,0.75) {Line\\failures\\$(\scenarioIdx)$};
	\draw [thick,->] (3.75,1.7) -- (3.75,0.15);
	\node[align=center] at (3.75,2.1) {Storm\\passing};
	\node[align=center] at (5.5,1) {Line repairs\\($\yc{}{}$)};
	\node[align=center,font=\fontsize{7}{7}\selectfont] at (5.5,-1) {($\xc{}{}$)\\Dispatch\\within\\ microgrids};
	\draw [thick,->] (3.75,-0.85) -- (3.75,-0.15);
	\node[align=center] at (3.75,-1.25) {$\time=\time_2$\\$\period=0$};
	\draw [thick,->] (7.25,-0.85) -- (7.25,-0.15);
	\node[align=center] at (7.25,-1.25) {\\$\period=\nperiod$};
	\node[align=center] at (7.25,2.1) {Reconnect\\to grid};
	\draw [thick,->] (7.25,1.7) -- (7.25,0.15);
	\draw [thick,->] (0,0) -- (8,0);
	\draw [thick,decorate,decoration={brace,amplitude=6pt,raise=0pt}] (3.9,0.2) -- (7.1,0.2);
	\draw [thick,decorate,decoration={brace,amplitude=6pt,raise=0pt}] (7.1,-0.2) -- (3.9,-0.2);
	\foreach \x in {-1,...,13}
	\draw [thick,-] (4+0.25*\x,0.1) -- (4+0.25*\x,-0.1);
	\end{tikzpicture}
	\caption{Timeline of events and decision stages. The DER placement decision ($\first$) is made before the storm hits the network ($t=\time_1$). Uncertainty $\scenarioIdx$ is realized over the course of the storm. After passing of the storm ($t=\time_2$, $\protect\period = 0$), optimal power flow and line repair decisions ($x,y$) are made. At $\protect {\period=\nperiod}$ (end of repair time horizon), the network is fully restored. } 
	\label{fig:timeline}
	\vspace{-3mm}
\end{figure}

Our two-stage stochastic mixed-integer problem considers the DER placement decisions in Stage I (pre-storm), and a multi-period repair problem with DER dispatch within each microgrid in Stage II (post-storm). The objective is to minimize the sum of the cost incurred in DER allocation and the expected cost of unmet demand during the time period of repair and recovery operations. For a given DER allocation (placement) and a realization of DN component disruptions, Stage II is a multi-period problem in which line repair schedules and dispatch within each microgrid are jointly determined. From a practical viewpoint, each period can be viewed as one work shift of the repair crews and the number of repairs per period is constrained. In the $0^\text{th}$ period, the subnetworks formed as a result of disruptions start to operate as microgrids using the available DER supply. In the subsequent time periods, damaged lines are repaired, permitting connections between smaller microgrids to progressively form larger microgrids. In the last time period, the DN is connected back to the main grid, and normal operation is restored. Crucially, the Stage II problem relies on an estimate of the total number of time periods needed for full recovery. We also utilize a novel model of linear power flow for a microgrid island with parallel operation of multiple DER inverters. \Cref{fig:timeline} summarizes the order of events and decisions in our formulation. 

Our formulation considers a tree DN with nodes and distribution lines $\G=(\N\bigcup\{0\},\E)$, where $\N$ denotes the set of all DN nodes. The substation node is labeled as 0, and it also forms the connection to the bulk supply through a transmission network. The set $\E$ denotes the set of directed edges, such that the edges are directed away from the substation node. The first-stage problem is as follows \cite{Shapiro}: 
\begin{align}\label{eq:sooFirstStage}
\begin{aligned}
\min_{\first\in\First} \left\{\soo\left(\first\right) \coloneqq \costFirstStage^T \first + \mathbb{E}_{ \uncertaintyR \sim \probFails} \recourse\left( \first, \uncertaintyR\right) \right\},
\end{aligned}
\end{align}
where $\first$ denotes a resource allocation strategy to be chosen from the set of feasible strategies $\First$. The uncertainty in the random vector $\uncertaintyR$ characterizes the random failures of distribution lines and has a probability distribution $\probFails$ defined over the set of possible line failure scenarios $\setScenarios \coloneqq \{0,1\}^{\E}$. $\costFirstStage$ is a length-$|\N|$ vector of the allocation cost per unit resource at the nodes, $\costFirstStage^T\first$ is the cost of resource allocation and $\mathbb{E}_{\uncertaintyR\sim\probFails}\recourse(\first,\uncertaintyR)$ is the expected cost of unmet demand under allocation scheme $\first$. 

To model the post-storm multi-period dispatch with repair scheduling, we consider an a priori fixed time horizon with $\nperiod$ periods. Let $\M \coloneqq \{0,1,\cdots,\nperiod\}$ denote the set of all periods. We denote a period by $\period$. For a specific realization of line failures $\scenarioIdx\in\setScenarios$, $\recourse(\first,\scenarioIdx)$ denotes the optimal value of the second-stage problem which is given as follows: 
\begin{align}\label{eq:sooSecondStage}
\begin{aligned}\small
\recourse\left(\first, \scenarioIdx\right) := &\min_{\xc{\scenarioIdx}{},\yc{\scenarioIdx}{}} &&
\ssum_{\period = 0}^{\nperiod}\ \costSecondStage^T\xc{\ts}{} \\
& \text{ s.t.} &&\yc{\scenarioIdx}{}\in \Yc{}{}\left(\scenarioIdx\right), \quad  \xc{\scenarioIdx}{} \in \Xc{}{}\left(\first,\scenarioIdx,\yc{\scenarioIdx}{}\right),
\end{aligned}
\end{align}
where the scenario-specific second-stage decision variables $\xc{\scenarioIdx}{} = \{\xc{\ts}{}\}_{\period\in\M}$ and $\yc{\scenarioIdx}{} = \{\yc{\ts}{}\}_{\period\in\M}$ respectively denote the collection of dispatch and line repair actions for each period. For a failure scenario $\scenarioIdx\in\setScenarios$, $\Yc{}{}\left(\scenarioIdx\right)$ denotes the set of feasible repair schedules, and $\Xc{}{}\left(\first,\scenarioIdx,\yc{\scenarioIdx}{}\right)$ denotes the set of feasible power flows under the DER allocation $\first$ and chosen line repair schedule $\yc{\scenarioIdx}{}\in\Yc{}{}\left(\scenarioIdx\right)$. 
The number of all possible scenarios grows exponentially ($2^{|\N|}$) for a network with $|\N|$ number of nodes, such that enumeration of scenarios in $\setScenarios$ is not feasible for sufficiently large $|\N|$. This challenge renders calculating $\mathbb{E}_{\uncertaintyR\sim\probFails}\recourse(\first, \uncertaintyR)$ computationally intractable for large networks. Using the sample average approximation (SAA) method~\cite{Shapiro}, one can obtain an approximate solution to the stochastic optimization problem \eqref{eq:sooFirstStage} by solving the following mixed integer program (MIP):

\begin{equation}\label{eq:sooEmpirical}
\begin{aligned}
&\min_{\first\in\First}  \left\{
\hat \soo_{\setScenariosSmall}(\first):= \costFirstStage^T\first + \frac{1}{\nscenario}\ssum_{\scenarioIdx \in \setScenariosSmall}  \recourse(\first, \scenarioIdx) \right\},
\end{aligned}
\end{equation}
where $\setScenariosSmall \subset \setScenarios$ is a suitably chosen (preferably small) subset of the set of failure scenarios, $\nscenario \coloneqq |\setScenariosSmall|$, and $\hat \soo_{\setScenariosSmall}(\first)$ is the SAA objective value of the first-stage problem obtained using $\nscenario$ samples drawn from the distribution $\probFails$. By solving the MIP, we obtain the first-stage solution $\first$ as well as the second-stage solutions $\xc{\scenarioIdx}{}$ and $\yc{\scenarioIdx}{}$ given by $\recourse(\first, \scenarioIdx)$. The set of constraints for Eq. \ref{eq:sooEmpirical} will be discussed in \cref{sec:physicalNetwork}.

The outline of the paper is as follows. In \cref{sec:lineFailureModel}, we describe the storm wind field prediction and NHPP failure model. In \cref{sec:physicalNetwork}, we describe the DER placement model, repair scheduling model, DER dispatch model, and \textit{LinDistFlow} model for islanded microgrids. In \cref{sec:computational}, we describe our computational results on a 12-node network. Finally, we mention some future extensions to our work in \cref{sec:conludingRemarks}.

\section{Failure model}\label{sec:lineFailureModel}
In general, the probability distribution $\probFails$ governing the random vector of line failures $\scenario$ in the first-stage problem (Eq. \ref{eq:sooFirstStage}) can be supported over $\{0,1\}^{\E}$. For some $\scenario= \scenarioIdx$,  we use the convention $\scenarioIdx_e=1$ to denote that line $e$ has failed, and 0 otherwise. To capture the physical impact of the storm wind field on DN components, we adopt an approach that characterizes $\probFails$ by combining: (i) a wind velocity prediction model given a forecast of the storm track and (ii) a Non-Homogeneous Poisson Process (NHPP) model for prediction of line failure rates. The two-dimensional area of the power network is divided into grids of size $\sim$1km$\times$1km, which are indexed by $\grid$ and form the set $\gridSet$. We estimate the wind velocity and Poisson failure rate within each grid every hour while the storm is passing over the network. The failure probabilities of distribution lines, which are roughly of length 1km in our study and can pass through multiple grids, are a function of the Poisson rates. 

The prediction of velocity measurements $\vel{}{h,t}$ in each grid $\grid$ and time $\time\in[\time_1,\time_2]$ is based on the storm center location at time $\time$ and three wind field parameters: maximum intensity $\Vm$, radius of maximum winds $\Rm$, and shape parameter $\B$. The predicted velocities can be obtained from the classical Holland model which expresses velocity $\vel{}{\grid,\tstorm{}}$ as a function of distance $\radius{}{\grid,\tstorm{}}$ from the storm center \cite{Holland}:
\begin{equation}\small
\label{eq:Holland}
\begin{aligned}
\vel{}{\grid,\tstorm{}}=\Vm\left(\frac{\Rm}{\radius{}{\grid,\tstorm{}}}\right)^{\B/2}\exp\left(1-\left(\frac{\Rm}{\radius{}{\grid,\tstorm{}}}\right)^{\B}\right)^{1/2}.
\end{aligned}
\end{equation}

An estimate of location and time-dependent Poisson failure rates $\PPI{}{\grid,\time}$, per unit time (hr) and line length (km), can be obtained using a quadratic NHPP model \cite{Zhou, AS, LiGengfeng}:
\begin{equation}
\label{eq:PPI}
\PPI{}{\grid,\time} =
\begin{cases}
\Big(1+\NHPPscale\Big(\left(\frac{\vel{}{\grid,\time}}{\velCrit}\right)^2-1\Big)\Big)\PPInorm, & \text{if }\vel{}{h,t}\geq \velCrit\\
\PPInorm, & \text{if }\vel{}{h,t}<\velCrit.
\end{cases}
\end{equation}
The failure rate is $\PPInorm$ if $\vel{}{\grid,\time}$ is below a critical velocity $\velCrit$ and increases quadratically with respect to wind speed above $\velCrit$. We use $\NHPPscale=4175.6$, $\velCrit=20.6$ m/s, and $\PPInorm=3.5\times 10^{-5}$ failures/hr/km. These parameters were estimated in a previous study using historical storm data that includes Category 1-3 storms \cite{LiGengfeng}. \footnote{Previous studies applying the quadratic NHPP model simulate the total number of line failures within a region while assuming a spatially-constant velocity in the region at every time step.} 

The cumulative intensity function $\CDF{}{\grid}$ per km at grid $h$ from storm arrival ($\tstormS$) over the network to its departure ($\tstormF$) is obtained by integrating the rate over the time interval:
\begin{equation}
\begin{aligned}
\CDF{}{\grid}(\time_2-\time_1)=\int_{\time_1}^{\time_2} \PPI{}{\grid,\time}d\time{}{}.
\end{aligned}
\end{equation}
Because the failure rate is given at hourly intervals and measured per hour, $\CDF{}{h}(\cdot)$ is approximately calculated as
\begin{equation}
\begin{aligned}
\CDF{}{\grid}(\time_2-\time_1)=\ssum_{\time=\time_1}^{\time_2} \PPI{}{\grid,\time}.
\end{aligned}
\end{equation}
Recall that a power line may span multiple grids. Let $\lineLength{}{e,\grid}$ denote the length of edge $e$ in grid $\grid$. The cumulative intensity function for line $e$ is 
\begin{equation}
\begin{aligned}
\CDFEdge{}{e}(\time_2-\time_1)=\ssum_{\grid\in\gridSet}\lineLength{}{e,\grid}\CDF{}{\grid}(\time_2-\time_1).
\label{eq:lineCDF}
\end{aligned}
\end{equation}
In our model, a line that has failed will remain failed for the remaining duration of the storm, such that a line can either never fail or fail once. The failure probability of line $e$ over the storm interval $\time_2-\time_1$ is
\begin{equation}
\begin{aligned}
\probFail{}{e}(\time_2-\time_1)=1-e^{-\CDFEdge{}{e}(\time_2-\time_1)}.
\label{eq:probLineFail}
\end{aligned}
\end{equation} 
Finally, the probability of scenario $\scenarioIdx$ is given by:
\begin{equation}\small
\begin{aligned}
\hspace{-0.1cm}\probExp(\scenarioIdx)=\pprod_{e\in\E[]}\big(\scenarioIdx_e\probFail{}{e}(\time_2-\time_1)+(1-\scenarioIdx_e)(1-\probFail{}{e}(\time_2-\time_1)\big)
\end{aligned}
\label{eq:probScenario}
\end{equation}
which characterizes the failure probability distribution $\probFails$. \footnote{In fact, location-specific features such as soil properties and vegetation can also impact the failure probabilities \cite{Whipple}. The effects of these inputs on the failure rates can be included by training a Poisson regression model, provided such data is available.}  

The SAA method relies on solving the two-stage problem for a subset of scenarios $\setScenariosSmall$. To select $\setScenariosSmall$, we begin by generating 1,000 realizations $\scenarioIdx$ of the random vector $\scenario$ and sort $\scenarioIdx$ in decreasing order of their probabilities $\probExp(\scenarioIdx)$. Then, we randomly choose a small subset  $\setScenariosSmall$ ($\le 10$ number of scenarios) from the 100 most probable of these scenarios. This procedure returns a subset that is representative of the most likely failure scenarios in the DN. The question of how well this sample represents the distribution $\probFails$ will be addressed in our future work. Nevertheless, we use the set  $\setScenariosSmall$ as an input to the problem \eqref{eq:sooEmpirical}, the variables and constraints of which are described in the next section. 

\section{Problem formulation}\label{sec:physicalNetwork}
We first define notation related to DN line parameters. A distribution line $e\in\E$ connects a  node $j$ to its parent node $i$ in the tree network. Here $i$ and $j$ are the from and to nodes of line $e$, which are denoted by $\fromNode{e}$ and $\toNode{e}$, respectively. Each line has a complex impedance $\impedance{e}=\resistance{e}+\j\reactance{e}$ where $\resistance{e}>0$ and $\reactance{e}>0$ denote the resistance and inductance of the line $e$, respectively, and $\j=\sqrt{-1}$. Also, let $\NN:=|\N|$. Now, we can detail our formulation for problem \eqref{eq:sooEmpirical}.

\subsection{DER placement model}
Recall that the first-stage decision of problem \eqref{eq:sooEmpirical} is the placement of DERs at DN nodes. Let $\setSite\subseteq \N$ denote the subset of nodes where DER sites can be developed. Let $\ysc{}{}\in{\{0,1\}}^{\setSite}$ be a vector, where $\ysc{}{i} = 1$ if a DER site is developed at node $i$ and  $\ysc{}{i} = 0$ otherwise. Let $\setDER$ denote the set of available DERs. The ratings of such DERs (e.g. synchronous generators that run on diesel or natural gas, or photovoltaic generators~\cite{derSiteArticle,derSiteArticle2,derSolarHurricane}) vary from 1 kilowatt to several megawatts~\cite{derSiteFEMAArticle}, so placement of DERs by size is a relevant question. For simplicity, we assume in our study that the DERs are homogeneous, i.e. all have the same rating. 

The primary costs of DER placement pertain to site development (e.g. land acquisition or building enclosures to protect DERs from natural disasters); protection of fuel delivery systems that include tanks, pumps and pipelines for fuel; transportation of fuel; and installation of the DERs ~\cite{derSiteArticle,derSiteArticle2}. We assume that the cost of site development is significantly higher than other costs, and thus refer to $\costFirstStage$ as $\Csite$ in which `SD' denotes `site development'. Let $\Csite\in\R_+^\setSite$ be a cost vector such that $\Csite_i$ denotes the cost of developing a DER site  at node $i\in\setSite$. Depending on the size and location of the DER site, $\Csite_i$ is assumed to be between tens of thousands to a hundred thousand dollars. 

Let $\ygc{}{}\in\{0,1\}^{\setSite\times\setDER}$ denote a map of DER allocation to candidate sites, such that $\ygc{}{\site\der} = 1$ denotes that a DER $\der$ is allocated at site $\site$. A site $\site\in\setSite$ is operational if and only if there is at least one DER unit  allocated to that site, i.e.
\begin{subequations}\label{eq:siteConditions}
	\begin{alignat}{8}
	\label{eq:siteConditionNecessary}
	\ysc{}{i} &\le \ssum_{\der\in\setDER} \ygc{}{i\der} \qquad &&\forall \ \site \in\setSite\\
	\label{eq:siteConditionSufficient}
	\ygc{}{i\der} & \le \ysc{}{i}\qquad &&\forall \ \der\in\setDER, \site \in\setSite.
	\end{alignat}
\end{subequations}
Since there are at most $\abs{\setDER}$ DERs, \Cref{eq:siteConditionSufficient} can be rewritten with fewer constraints as $\ssum_{\der\in\setDER} \ygc{}{i\der} \le \ysc{}{i}\abs{\setDER} \ \forall \ \site \in\setSite.$

Clearly, a DER $\der$ can be allocated to at most one site, i.e.
\begin{equation}\label{eq:oneNodeCondition}
\ssum_{i\in\setSite} \ygc{}{i\der} \le 1\qquad \forall \ \der\in\setDER.
\end{equation}
In practice, the number of DERs are limited by supply. Let $\genBudget$ denote the maximum number of DERs that can be allocated in a DN. Then
\begin{equation}\label{eq:genResourceConstraint}
\ssum_{i\in\setSite} \ssum_{\der\in\setDER} \ygc{}{i\der} \le \genBudget. 
\end{equation}

Thus, the first stage decision variable in problem \eqref{eq:sooEmpirical} of joint site development and DER allocation decisions can be defined as $\first\coloneqq \left(\ysc{}{},\ygc{}{}\right)$. The set of feasible resource allocation strategies is $\First \coloneqq \{\left(\ysc{}{},\ygc{}{}\right) \in \setSite\times\setDER \ | \ \eqref{eq:siteConditions}-\eqref{eq:genResourceConstraint} \text{ hold} \}$. 

\subsection{Multi-period repair scheduling model}
Recall that the second-stage decision of problem \eqref{eq:sooEmpirical} is the multi-period line repair schedule and dispatch within microgrids. We assume that at period $\period = 0$, the uncertainty of line failures due to the storm is realized, and the pre-placed DERs can be  dispatched to supply power. From period $\period=1$, the utility crew starts repairing the damaged lines subject to resource constraints. We choose $\nperiod$ large enough to allow all line repairs in the DN to be complete. Furthermore, we consider that the system performance, in terms of value of demand met, can be fully restored after the DN is reconnected to the main grid. But in the worst-case scenario, this reconnection is likely to happen after the line repairs are completed at $\period=\nperiod$. Thus, we constrain the DN to be reconnected back to the main grid at period $\period=\nperiod$ for worst-case analysis ~(see \cref{fig:timeline}). 

Consider a scenario $\scenarioIdx\in\setScenariosSmall$ denoting the locations of line failures and let $\E[\scenarioIdx]$ denote the set of failed edges under scenario $\scenarioIdx$. For each scenario and $\period \in \M$, let $\ylinec{}{} \in {\{0,1\}}^{\E[s] \times \M}$ denote the decision variables concerning repair of failed lines, where $\ylinec{\ts}{e} = 1$ for $(e,\period)\in\E[s]\times \M$ denotes that the line $e$ is repaired during the $\period^{th}$ time period. The repair crew usually needs some time (typically 24-48 hours) for damage assessment and planning of restoration actions after the storm ends, and thus cannot start the repair of lines immediately after the storm. Hence, we assume that no lines are repaired in the first period, i.e. 
\begin{equation}\label{eq:norepairFirstPeriod}
\ylinec{\period,\scenarioIdx}{e} = 0 \quad \forall\  e \in \E[s], \period = 0. 
\end{equation}
We also assume that at most $\crewBudget$ number of lines can be repaired during any time period, i.e. 
\begin{alignat}{6}
\label{eq:lineRepairCrewConstraint}
\ssum_{e\in\E[s]}\ \ylinec{\period,\scenarioIdx}{e} &\le \crewBudget \qquad &&\forall\ \period\in\M.
\end{alignat}
Here $\crewBudget$ models the combined capability of the crew personnel and equipment in repairing lines in a single period. \footnote{It is indeed possible that the maximum number of repairable lines may vary across periods, which are realistically about a day long. For the sake of simplicity, we chose a fixed $\crewBudget$.} Finally, we constrain that the DN is connected to the main grid at period $\period=\nperiod$, i.e.
\begin{equation}\label{eq:mainGridConnected}
\ylinec{\period,\scenarioIdx}{e} = 1, \quad\period=\nperiod, e \in\E, \fromNode{e} = 0, 
\end{equation}
where $\fromNode{e} = 0$ denotes that line $e$ connects the DN to the substation node $0$. 

For each scenario and $\period \in \M$, let $\klinec{}{}\in\{0,1\}^{\E \times \M}$ denote the variables regarding whether the line $e\in\E[]$ is operational at time $\period\in\M$, where $\klinec{\period,\scenarioIdx}{e} = 1$ if the line $e$ is not operational at period $\period$ and $\klinec{\period,\scenarioIdx}{e} = 0$ otherwise. We consider that the lines not damaged during the storm remain operational during all periods: 
\begin{equation}\label{eq:undamagedLines}
\klinec{\ts}{e} = 0\quad \forall\ \period\in \M, e \notin\E[s].
\end{equation}

The connectivity of lines damaged by the storm at period $\period = 0$ is as follows:
\begin{equation}\label{eq:initialLineDamage}
\klinec{\period,\scenarioIdx}{e} = 1, \qquad\forall\ e \in\E[s],  \period  = 0.
\end{equation}
To model the disconnect between the DN and substation until $\period=\nperiod$, we set $kl^{0,s}_{e} = 1$ for $e\in\E[s]$ such that $\fromNode{e} = 0$.

A failed line becomes operational after it is repaired, and thereafter continues to remain operational, i.e.
\begin{equation}\label{eq:lineRepairChange}
\klinec{\period,\scenarioIdx}{e} = \klinec{\period-1,\scenarioIdx}{e} - \ylinec{\period,\scenarioIdx}{e} \quad\forall\ \period \in \M\setminus 0, e\in\E[s].
\end{equation}
Since, for all $\period$, the variables $\klinec{\period,\scenarioIdx}{e}$ and $\ylinec{\period,\scenarioIdx}{e}$ can only take binary values,  \eqref{eq:lineRepairChange} automatically ensures that a failed line can at most be repaired once, i.e. 
$\forall\ e\in\E[s], \ \sum_{\period = 0}^{\nperiod}\ylinec{\period,\scenarioIdx}{e} \le 1$. 

Thus, the repair schedule variable for each scenario $\scenarioIdx$ in \cref{eq:sooSecondStage} can be denoted as $\yc{\scenarioIdx}{} \coloneqq \{(\ylinec{\ts}{e}, \klinec{\ts}{e})\}_{e\in\E,\period\in\M}$, and the set of feasible repair schedules can be written as $\Yc{}{}\left(\scenarioIdx\right) \coloneqq \{\yc{}{} \in \E\times\M \ | \ \eqref{eq:norepairFirstPeriod}-\eqref{eq:lineRepairChange} \text{ holds} \}$. 

\subsection{Multi-period dispatch model}
As lines are repaired, the connectivity between the microgrid islands is restored, and loads are gradually reconnected. Thus, the DERs need to be redispatched after each period to meet the demand of reconnected loads. Below we present a multi-period dispatch model of DERs. Henceforth, we drop the notation $\forall\ \period\in\M, \scenarioIdx \in\setScenariosSmall$ for the sake of brevity. 

\paragraph*{DER model} 
Let $\pgc{\ts}{i\der}$ and $\qgc{\ts}{i\der}$ respectively denote the real and reactive power generated by DER $\der$ at node $i$. Depending on whether a DER $\der$ is allocated at site $i\in\setSite$, its contribution to the power generated at node $i$ at any time $t$ is constrained as:
\begin{align}\label{eq:derConnectedConstraint}
\begin{aligned}
0 &\le &&\pgc{\ts}{i\der} &&\le \ygc{}{i\der} \pgc{max}{\der} && \forall\ i \in \setSite, \der \in \setDER\\
& &&\abs{\qgc{\ts}{i\der}} &&\le \pfc{max}{} \pgc{\ts}{i\der} && \forall\ i \in \setSite, \der \in \setDER
\end{aligned}
\end{align}
where $\pfc{max}{}$ denotes the $\tan\arccos$ of maximum power factor.
The active and reactive power contributions of a DER to non-DER site nodes are zero, i.e.
\begin{equation}
\pgc{\ts}{i\der} = \qgc{\ts}{i\der} = 0 \quad \forall\ i\in\N\backslash\setSite, \der\in\setDER.
\end{equation} 

In a microgrid island, each DER can adjust its reactive power output depending on the voltage of the node to which it is allocated. This capability can be modeled by a voltage droop control equation:
\begin{equation}\label{eq:islandVoltDroop}
\begin{split}
\abs{\nuc{\ts}{i} - \left(\nuc{ref}{\der} - \kqc{\der}\qgc{\ts}{i\der}\right)} \le \left(1-\ygc{}{i\der}\right)\bigM\\
\forall\ i\in\setSite,\der\in\setDER, \period\in\{1,...,\nperiod-1\}
\end{split}
\end{equation}
where $\nuc{\ts}{i}$ denotes the voltage at node $i$; $\kqc{\der}$ denotes the voltage droop coefficient of the DER $\der$; $\nuc{ref}{\der}$ denotes the idle (no load) terminal voltage reference setpoint of the DER~\cite{droopControl}; and $\bigM$ is a large constant.
If $\ygc{}{i\der} = 1$, \eqref{eq:islandVoltDroop} simplifies to $\qgc{\ts}{i\der} = \frac{1}{\kqc{\der}}\left(\nuc{ref}{\der} - \nuc{\ts}{i}\right)$, which models the standard droop control equation that determines the reactive power contribution of the DER at node~$i$ to help voltage regulation of the islanded microgrid. \footnote{Turbine-based DERs such as diesel generators are AC power sources, so they can contribute to voltage regulation via in-built excitation systems that utilize automatic voltage regulators~\cite{derExcitationSystems}. In contrast, DERs such as batteries or photo-voltaic generators are DC power sources, which are connected to the DN via inverters.  Such DERs can contribute to voltage regulation if their inverters are set in the voltage source inverter (VSI) control mode~\cite{islandingControlStrategies}.} Once the DN is connected to the main grid, the \enquote{stiff} AC system of the bulk power grid determines the terminal voltage of the DERs. Hence, the voltage droop equation does not apply at period $\period = \nperiod$. \footnote{In our model, each DER within an island contributes to the voltage regulation of the microgrid island, simultaneously with other DERs in the same island~\cite{parallelInverterDroopControl}. A suitable extension of our model can also allow some of the DERs to operate in the PQ control mode, in which the DERs' active and reactive power supply is fixed. Moreover, since the DERs can also contribute to frequency regulation of the islanded microgrid, our model can also be extended to consider the frequency droop control equations.}

Next, we describe the load model. We  consider only the constant power model for loads. 
\paragraph*{Load model}
Let $\pcc{max}{i}+\j\qcc{max}{i}$ denote the nominal power demand at node~$i$; $\kcc{\ts}{i}\in\{0,1\}$  the load connectivity where $\kcc{\ts}{i} = 1$ if the load at node $i$ is disconnected at time period $\period$. When connected, a load's actual consumption can be scaled down to a fraction of its nominal demand, particularly if the available power supply from DERs is not sufficient to meet nominal demand. We model such flexibility by introducing a load control parameter  $\lcc{\ts}{i}\in [\lcc{min}{i},1]$ (when $\kcc{\ts}{i} = 0$, otherwise $\lcc{\ts}{i}=0$), where   $\lcc{min}{i} \in [0,1]$ denotes the minimum fraction of the load's nominal demand that should be satisfied provided the load is connected, i.e.
\begin{align}\label{eq:loadControlEquation} 
\begin{aligned}
\pcc{\ts}{i} &=  \lcc{\ts}{i}\ \pcc{max}{i} \quad \forall\ i \in \N \\			
\qcc{\ts}{i} &=  \lcc{\ts}{i}\ \qcc{max}{i} \quad \forall\ i \in \N,
\end{aligned}
\end{align}
where $\pcc{\ts}{i}$ and $\qcc{\ts}{i}$ respectively denote real and reactive power consumed at node $i$, and
\begin{equation}\label{eq:loadControlParameterConstraint}
\left(1-\kcc{\ts}{i}\right)\lcc{min}{i} \le  \lcc{\ts}{i} \le  \left(1-\kcc{\ts}{i}\right) \qquad \forall \ i\in\N.
\end{equation}
The load connectivity depends on whether the nodal voltage lies within safe operating bounds, i.e.
\begin{align}\label{eq:voltageDisconnectLoads}		
\begin{aligned}
\kcc{\ts}{i} &\ge \nucc{min}{i} - \nuc{\ts}{i} \quad && \forall\ i\in \N\\
\kcc{\ts}{i} &\ge \nuc{\ts}{i} - \nucc{max}{i} \quad && \forall\ i\in \N,
\end{aligned}
\end{align}
where $\nucc{min}{i}$ and $\nucc{max}{i}$ denote the lower and upper voltage bounds for the loads at node $i$. 

The net actual real and reactive power consumed (denoted by $\ptc{\ts}{i}$ and $\qtc{\ts}{i}$, respectively) at node $i$ is given by:
\begin{align}
\begin{aligned}
\ptc{\ts}{i} &= \pcc{\ts}{i} -  \ssum_{\der\in\setDER}\pgc{\ts}{i\der}  \qquad \forall\ i\in\N\\
\qtc{\ts}{i} &= \qcc{\ts}{i} -  \ssum_{\der\in\setDER}\qgc{\ts}{i\der} \qquad  \forall\ i\in\N.
\end{aligned}\label{eq:totalPowerConsumption}
\end{align}

\subsection{Linear Power Flow model for microgrids} 
We now develop a novel linear power flow model for microgrids based on the classical \textit{LinDistFlow} model. The linearity of these equations provides a computational advantage over the nonlinear power flow model. 

We can write the standard power conservation equations for real and reactive power flows as follows:
\begin{align}\label{eq:powerConservation}
\begin{aligned}
\Pc{\ts}{e} &= \displaystyle\sum_{l: \fromNode{l} = \toNode{e}} \Pc{\ts}{l} + \ptc{\ts}{j}, && \forall\ e \in \E, j = \toNode{e} \\
\Qc{\ts}{e} &= \displaystyle\sum_{l: \fromNode{l} = \toNode{e}}\Qc{\ts}{l} + \qtc{\ts}{j}, && \forall\ e \in \E, j = \toNode{e}.
\end{aligned}
\end{align}

Since the failed lines are not operational, there are no power flows on these lines until they are repaired, i.e.
\begin{align}\label{eq:islandingCapacity}
\begin{aligned}
\abs{\Pc{\ts}{e}} & \le \left(1-\klinec{\ts}{e}\right)\bigM \qquad && \forall\ e\in\E\\
\abs{\Qc{\ts}{e}} & \le \left(1-\klinec{\ts}{e}\right)\bigM \qquad && \forall\ e\in\E. 
\end{aligned}
\end{align}
Similarly, the voltage drop equation of the \textit{LinDistFlow} model along a line $e$ is enforced only if line $e$ is operational: 
\begin{align}\label{eq:voltageDrop} 
\begin{aligned}
\abs{\nuc{\period,\scenarioIdx}{j} -  \left(\nuc{\period,\scenarioIdx}{i} -  2\left(\resistance{e}\Pc{\period,\scenarioIdx}{e} +  \reactance{e}\Qc{\period,\scenarioIdx}{e}\right)\right)} \le \bigM\klinec{\period,\scenarioIdx}{e} \\ \forall \ e \in \E, i = \fromNode{e}, j = \toNode{e}.
\end{aligned}
\end{align}
Note that if a line $e$ is operational (i.e. $\klinec{\ts}{e} = 0$), then \cref{eq:voltageDrop} simplifies to $\nuc{\ts}{j} = \nuc{\ts}{i} - 2\left(\resistance{e}\Pc{\ts}{e} +  \reactance{e}\Qc{\ts}{e}\right)$, which is the standard voltage drop equation of the \textit{LinDistFlow} model. When the DN is connected back to the main grid, the substation voltage is assumed to be the nominal voltage, i.e.  $\nuc{\nperiod,\scenarioIdx}{0} = \nuc{nom}{}$. 

We define the dispatch variable for scenario $s$ in \cref{eq:sooSecondStage} as $\xc{\scenarioIdx}{} \coloneqq  \{\pgc{\ts}{},\qgc{\ts}{},$ $\lcc{\ts}{},\kcc{\ts}{},\ptc{\ts}{},\qtc{\ts}{},\Pc{\ts}{},\Qc{\ts}{},$ $\nuc{\ts}{}\}_{\period\in\M}$, and the set of feasible dispatch decisions as $\Xc{}{}\left(\first,\scenarioIdx,\yc{}{}\right) \coloneqq \{\xc{}{}\ |\  \eqref{eq:initialLineDamage}-\eqref{eq:voltageDrop} \text{ hold}\}$.
The SAA problem \eqref{eq:sooEmpirical} can be specifically written as: 
\begin{alignat}{8}\small
\small\hspace{-0.1cm} \min_{\first,\xc{}{},\yc{}{}} 
& \ \Bigg\{\sum_{i\in\setSite} \Csite_i \ysc{}{i} + 
\frac{1}{\nscenario}\sum_{\period = 0}^{\nperiod}\sum_{\node\in\N} [\Cload_i(1-\lcc{\ts}{i})+ \Cshed_i\kcc{\ts}{i}] \nonumber\Bigg\}\\
\small	 \text{s.t.  }&\ \first=(\ysc{}{},\ygc{}{}), \yc{\scenarioIdx}{}\in\Yc{}{}\left(\scenarioIdx\right), \xc{\scenarioIdx}{} \in \Xc{}{}\left(\first,\scenarioIdx,\yc{}{}\right) \  \forall \ \scenarioIdx\in\setScenariosSmall,
\end{alignat}
where $\Cload_i$ and $\Cshed_i$ are respectively the costs of load control and shedding at node $i$; $\xc{}{} \coloneqq \{\xc{\scenarioIdx}{}\}_{\scenarioIdx\in\setScenariosSmall}$; and $\yc{}{} \coloneqq \{\yc{\scenarioIdx}{}\}_{\scenarioIdx\in\setScenariosSmall}$. 

\subsection{Illustrative example}
\begin{figure}[htbp!]
	\centering
	\tikzset{every node/.append ={font size=tiny}} 
	\tikzstyle{dnnode}=[draw,circle, minimum size=0.5pt, inner sep = 2]
	\tikzstyle{dnedge}=[-, line width=1pt]
	\tikzstyle{vuledge}=[-, line width=3pt, red!70]
	\tikzstyle{swind}=[->, line width=2pt, blue!25, >=latex']
	\tikzstyle{dernode}=[circle, fill=blue, minimum size=0.5pt, inner sep = 2]
	\tikzstyle{blackoutnode}=[circle, fill=black, minimum size=0.5pt, inner sep = 2]
	\tikzstyle{graynode}=[draw=black, pattern=north west lines, fill = black, circle,  minimum size=0.5pt, inner sep = 2, fill opacity=0.2]
	\tikzstyle{graynodes}=[circle, pattern=north west lines, fill = black, minimum size=0.5pt, inner sep = 2,fill opacity = 0.6]
	\tikzstyle{failededge}=[-, densely dotted]
	\def \drawgrid {\draw[step=1,gray, ultra thin, draw opacity = 0.5] (0,0) grid (3,4);}
	\def \drawSubstation {\draw[-, line width = 2pt] (0.8,4.05) -- (2.2,4.05)  node [midway,above] {};}
	\def \drawZero {\node[dnnode] (0) at (1.5,3.85) {};}
	\def \drawA {\node[dnnode] (A) at (1.5,3.0) {};}
	\def \drawAone {\node[dnnode] (A1) at (2,2.5) {};}
	\def \drawAtwo {		\node[dnnode] (A2) at (2.75,2.75) {};}
	\def \drawB {\node[dnnode] (B) at (0.25,2.75) {} ;}
	\def \drawC {\node[dnnode] (C) at (0.25,1.5) {} ;}
	\def \drawD {\node[dnnode] (D) at (1.5,1.25) {} ;}
	\def \drawDone {\node[dnnode] (D1) at (1.5,0.25) {} ;}
	\def \drawDtwo {\node[dnnode] (D2) at (0.5,0.25) {} ;}
	\def \drawDthree {\node[dnnode] (D3) at (1.5,2.00) {} ;}
	\def \drawE {\node[dnnode] (E) at (2.75,1.25) {} ;}
	\def \drawF {\node[dnnode] (F) at (2.75,0.50) {} ;}
	\subfloat[]{\label{subfig:nominal}
		\begin{tikzpicture}[scale=0.6,every node/.append ={font size=tiny}]
		
		\drawgrid \drawSubstation
		
		\drawZero \drawA \drawAone \drawAtwo \drawB \drawC \drawD \drawDone \drawDtwo \drawDthree \drawE \drawF 
		
		\foreach \from/\to in {0/A, A/B, A/A1, A1/A2, B/C, C/D, D/E, D/D1, D1/D2, D/D3, E/F}
		\draw[dnedge] (\from) -- (\to);		
		
		\node at (0) {\tiny $\text{0}$};
		\node at (A) {\tiny $\text{A}$};
		\node at (A1) {\tiny $\text{G}$};
		\node at (A2) {\tiny $\text{H}$};
		\node at (B) {\tiny $\text{B}$};
		\node at (C) {\tiny $\text{C}$};
		\node at (D) {\tiny $\text{D}$};
		\node at (D1) {\tiny $\text{I}$};
		\node at (D2) {\tiny $\text{J}$};
		\node at (D3) {\tiny $\text{K}$};
		\node at (E) {\tiny $\text{E}$};
		\node at (F) {\tiny $\text{F}$};
		\end{tikzpicture}
	}
	\subfloat[]{\label{subfig:predictions}
		\begin{tikzpicture}[scale=0.6]
		\drawgrid \drawSubstation
		
		\drawZero \drawA \drawAone \drawAtwo \drawB \drawC \drawD \drawDone \drawDtwo \drawDthree \drawE \drawF 
		
		\node[dernode] (Dder) at (D) {} ;
		
		\foreach \from/\to in {0/A, A/B, A/A1, A1/A2, B/C, C/D, D/E, D/D1, D1/D2, D/D3, E/F}
		\draw[dnedge] (\from) -- (\to);		
		\foreach \from/\to in {0/A, B/C, D/E, D/D1, D/D3}
		\draw[vuledge] (\from) -- (\to);
		
		\foreach \mx/\my in {0.5/2.5, 2/3, 0.75/1.75, 2/2, 0.75/0.5, 2/0.75}	
		\draw[swind] (\mx,\my) to  (\mx+0.75,\my+0.25);
		\end{tikzpicture}
	}
	\subfloat[$\period = 0$]{\label{subfig:failures}
		\begin{tikzpicture}[scale=0.6]
		\drawgrid \drawSubstation
		
		\drawZero \drawA \drawAone \drawAtwo \drawB \drawC \drawD \drawDone \drawDtwo \drawDthree \drawE \drawF 
		
		\node[blackoutnode] (0) at (0) {};
		\node[blackoutnode] (A) at (A) {};
		\node[blackoutnode] (A1) at (A1) {};
		\node[blackoutnode] (A2) at (A2) {};
		\node[blackoutnode] (B) at (B) {} ;
		\node[dnnode] (C) at (C) {} ;
		\node[dnnode] (D) at (D) {} ;
		\node[blackoutnode] (D1) at (D1) {} ;
		\node[blackoutnode] (D2) at (D2) {} ;
		\node[blackoutnode] (D3) at (D3) {} ;
		\node[dernode] (Dder) at (D) {} ;
		\node[blackoutnode] (E) at (E) {} ;
		\node[blackoutnode] (F) at (F) {} ;
		
		\foreach \from/\to in { A/B, A/A1, A1/A2, C/D, D1/D2, E/F}
		\draw[dnedge] (\from) -- (\to);		
		\foreach \from/\to in {0/A, B/C, D/E, D/D1, D/D3}
		\draw[failededge] (\from) -- (\to);
		\end{tikzpicture}
	}
	\\	\subfloat[$\period = 1$]{\label{subfig:firstRepairs}
		\begin{tikzpicture}[scale=0.6]
		
		\drawgrid \drawSubstation
		
		\node[blackoutnode] (0) at (0) {};
		\node[blackoutnode] (A) at (A) {};
		\node[blackoutnode] (A1) at (A1) {};
		\node[blackoutnode] (A2) at (A2) {};
		\node[blackoutnode] (B) at (B) {} ;
		\node[graynode] (C) at (C) {} ;
		\node[dnnode] (D) at (D) {} ;
		\node[graynode] (D1) at (D1) {} ;
		\node[graynode] (D2) at (D2) {} ;
		\node[blackoutnode] (D3) at (D3) {} ;
		\node[dernode] (Dder) at (D) {} ;
		\node[graynode] (E) at (E) {} ;
		\node[graynode] (F) at (F) {} ;
		
		\foreach \from/\to in { A/B, A/A1, A1/A2, C/D, D/E, D/D1, D1/D2, E/F}
		\draw[dnedge] (\from) -- (\to);		
		\foreach \from/\to in {0/A,B/C, D/D3}
		\draw[failededge] (\from) -- (\to);
		
		\end{tikzpicture}
	}
	\subfloat[$\period = 2$]{\label{subfig:secondRepairs}
		\begin{tikzpicture}[scale=0.6]
		
		\drawgrid \drawSubstation
		\drawZero \drawA \drawAone \drawAtwo \drawB \drawC \drawD \drawDone \drawDtwo \drawDthree \drawE \drawF 
		
		\node[blackoutnode] (0b) at (0) {};
		\node[blackoutnode] (Ab) at (A) {};
		\node[blackoutnode] (A1b) at (A1) {};
		\node[blackoutnode] (A2b) at (A2) {};
		\node[graynodes] (Bb) at (B) {} ;
		\node[graynodes] (C) at (C) {} ;
		\node[dnnode] (D) at (D) {} ;
		\node[graynodes] (D1) at (D1) {} ;
		\node[graynodes] (D2) at (D2) {} ;
		\node[graynodes] (D3) at (D3) {} ;
		\node[dernode] (Dder) at (D) {} ;
		\node[graynodes] (E) at (E) {} ;
		\node[graynodes] (F) at (F) {} ;
		
		\foreach \from/\to in { A/B, A/A1, A1/A2, C/D, D/E, D/D1, D1/D2, E/F,B/C, D/D3}
		\draw[dnedge] (\from) -- (\to);		
		\foreach \from/\to in {0/A}
		\draw[failededge] (\from) -- (\to);
		
		\end{tikzpicture}
	}
	\subfloat[$\period = \nperiod$]{\label{subfig:reconnected}
		\begin{tikzpicture}[scale=0.6]
		\drawgrid \drawSubstation
		\drawZero \drawA \drawAone \drawAtwo \drawB \drawC \drawD \drawDone \drawDtwo \drawDthree \drawE \drawF 
		
		\foreach \from/\to in {0/A, A/B, A/A1, A1/A2, B/C, C/D, D/E, D/D1, D1/D2, D/D3, E/F}
		\draw[dnedge] (\from) -- (\to);		
		\foreach \from/\to in {0/A, B/C, D/E, D/D1, D/D3}
		\draw[dnedge] (\from) -- (\to);
		\end{tikzpicture}
	}
	\caption{The subfigures show (a) nominal DN (white nodes indicate no load control), (b) pre-storm DER allocation based on storm forecast (blue node denotes DER allocation, red lines illustrate a disruption scenario), (c) microgrid islands (dotted lines indicate failed lines, black nodes denote the loads that are completely unserved), (d) partial line repairs enable partial load restoration (light gray nodes), (e) line repairs completed leading to more load restoration, although with even more load control than before (dark gray nodes), and finally (f) reconnection to main grid and restoration of nominal performance. }
	\label{fig:Schematic}
\end{figure}
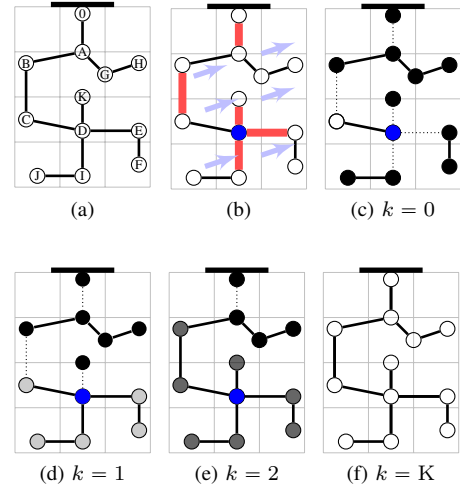

\Cref{fig:Schematic} provides an illustration of various aspects of our  formulation. In \Cref{subfig:nominal}, the DN is in nominal operating conditions, i.e. each node is connected to the grid and no load control is exercised. 

We obtain failure probabilities using \cref{eq:probLineFail} and generate $|\setScenariosSmall|$ failure scenarios. Suppose that the utility has resources to develop one DER site in Stage I. Further suppose that due to the powerflow and voltage drop constraints (\eqref{eq:powerConservation} and \eqref{eq:voltageDrop}), transmitting power across more than two lines results in voltage bound violations. Hence, from any site location, the DERs will be able to meet the demand of nodes which are at most two hops away from the site. For example, a DER placed at node H can only supply power to nodes A, G, and H, such that $\{$A,B,G,H$\}$ form the largest connected island. Hence, placing a site at node A can immediately serve demand of 4 loads. However, if we take into account the line repairs,  the maximum number of loads that can be reconnected to the DN (after repairs) is eight if the site is developed at node D. Thus, allocating the DER to node D is optimal because it maximizes number of loads served, though only two loads are served during the first period. 

In \Cref{subfig:failures}, we see that a failure scenario has been realized for period $\period = 0$ where the set of failed lines $\E[s]$ include lines (B,C), (D,E), (D,I), and (D,K). The loss of bulk power supply is represented by the DN's disconnection from the substation, i.e.  (0,A) $\in\E[s]$. Since the demand at just two nodes (C, D) is met in the first period, no load control is necessary (i.e. $\kcc{\ts}{i} = 0, \lcc{\ts}{i} = 1$ $\forall$ $i\in\{$C,D$\}$; and $\kcc{\ts}{i} = 1,\ \forall \ i \in \N \backslash \{$C,D$\}$).

Next, we schedule line repairs under the constraint $\crewBudget = 2$, i.e. at most two lines can be repaired in each period. Looking at the most number of loads that can be reconnected, the lines (D,E) and (D,I) should be repaired in the period $\period=1$ ($\ylinec{\ts}{e} = 1$ if $e\in\{$(D,E), (D,I)$\}$), and lines (D,K) and (B,C) should be repaired in the period $\period=2$  ($\ylinec{\ts}{e} = 1$ if $e\in\{$(D,K), (B,C)$\}$). Following this schedule, nodes E, F, I, J are connected while $\period=1$ (see~\Cref{subfig:firstRepairs}). The DER power supply is enough to serve only partial demand at these nodes ($\kcc{\ts}{i} = 0$ and $\lcc{\ts}{i} < 1$ for $i\in\{$C,D,E,F,I,J$\}$). The loads at nodes B, C and K are reconnected while $\period = 2$ due to further line repairs (see \Cref{subfig:secondRepairs}), and a greater portion of demand can be met ($\lcc{\period,s}{i} > \lcc{\period-1,s}{i}$ for $i\in\{$C,D,E,F,I,J$\}$). Due to power flow constraints, the loads at nodes A, G, and H cannot be reconnected until the complete network is restored. Finally, when the DN is reconnected to main grid ($\ylinec{\ts}{e} = 1$ for $e = $(0,A) and $\period = \nperiod$) the nominal operation of the DN is fully restored (see~\cref{subfig:reconnected}). 

\section{Evaluation}\label{sec:computational}

\subsection{Experimental setup}
We use a 12-node test feeder in our computational experiments. 6 randomly chosen nodes have one load each. For each homogeneous node $i$ with a load, $\Cshed_i =  1000$, $\Cload_i = 100$, $\lcc{min}{i} = 0.5$. The total capacity of available DERs is chosen to be 80\% of the total demand in the network. 

Next, we compute predictions for velocity fields $\vel{}{h,t}$ and Poisson failure rates $\PPI{}{\grid,\time}$ at every hour over one day, using the procedure described in \cref{sec:lineFailureModel}. To produce predictions of $\vel{}{h,t}$ using \eqref{eq:Holland}, we consider two different tracks of a Category 1 storm to account for expected uncertainty in the storm trajectory. The storm tracks we used (hereafter referred to as Track 1 and Track 2) differ primarily in that the storm eye wall (region of maximum winds) is farther away from the DN for Track 2, and thus the wind velocities in the DN are lower compared to the case with Track 1. We use synthetic values for the Holland parameters ($\Vm$, $\Rm$, $\B$) to produce predictions of $\vel{}{h,t}$ using \eqref{eq:Holland}. The velocity predictions $\vel{}{h,t}$ may be used to compute $\PPI{}{\grid,\time}$.

\begin{table}[!t]
	\renewcommand{\arraystretch}{1.3}
	\caption{Mean, minimum, and maximum failure probabilities of distribution lines (\textit{left side}) and median, minimum, and maximum island size (\textit{right side}) for the two tracks.}
	\label{tabFailures}
	\centering
	\begin{tabular}{|c||c|c|c||c|c|c|}
		\hline
		& \multicolumn{3}{c||}{Failure probability} & \multicolumn{3}{c|}{Size of islands} \\ \hline
		& Mean & Min & Max & Med. & Min & Max \\ \hline
		Track 1 & 0.63 & 0.56 & 0.75 & 1.58 & 1.08 & 4.20 \\ \hline
		Track 2 & 0.26 & 0.21 & 0.34 & 3.22 & 2.11 & 5.61  \\ \hline
	\end{tabular}
\end{table}

\begin{figure}[!htbp]
	\centering
	\subfloat{\includegraphics[width=0.5\textwidth, scale=0.12]{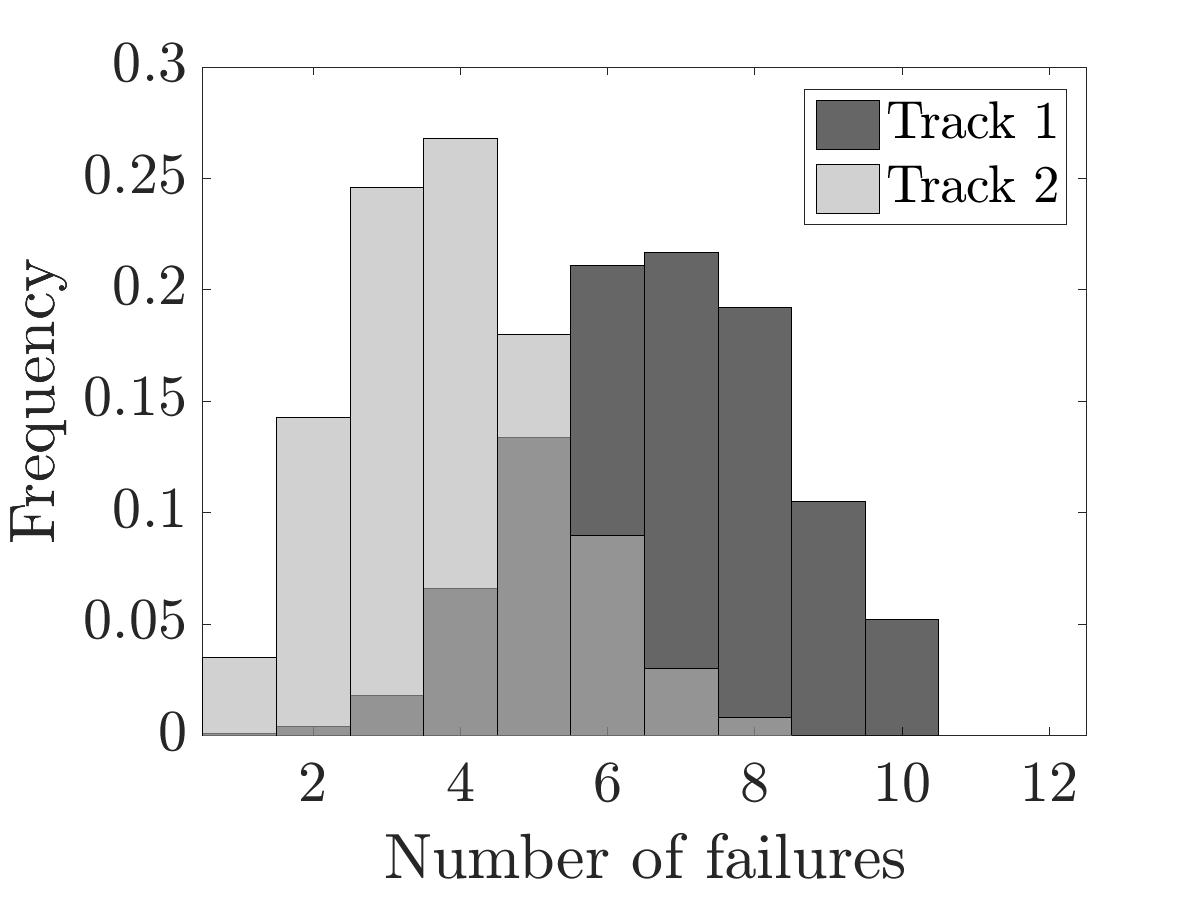}	\label{fig:SimFailures}}
	\subfloat{\includegraphics[width=0.5\textwidth, scale=0.12]{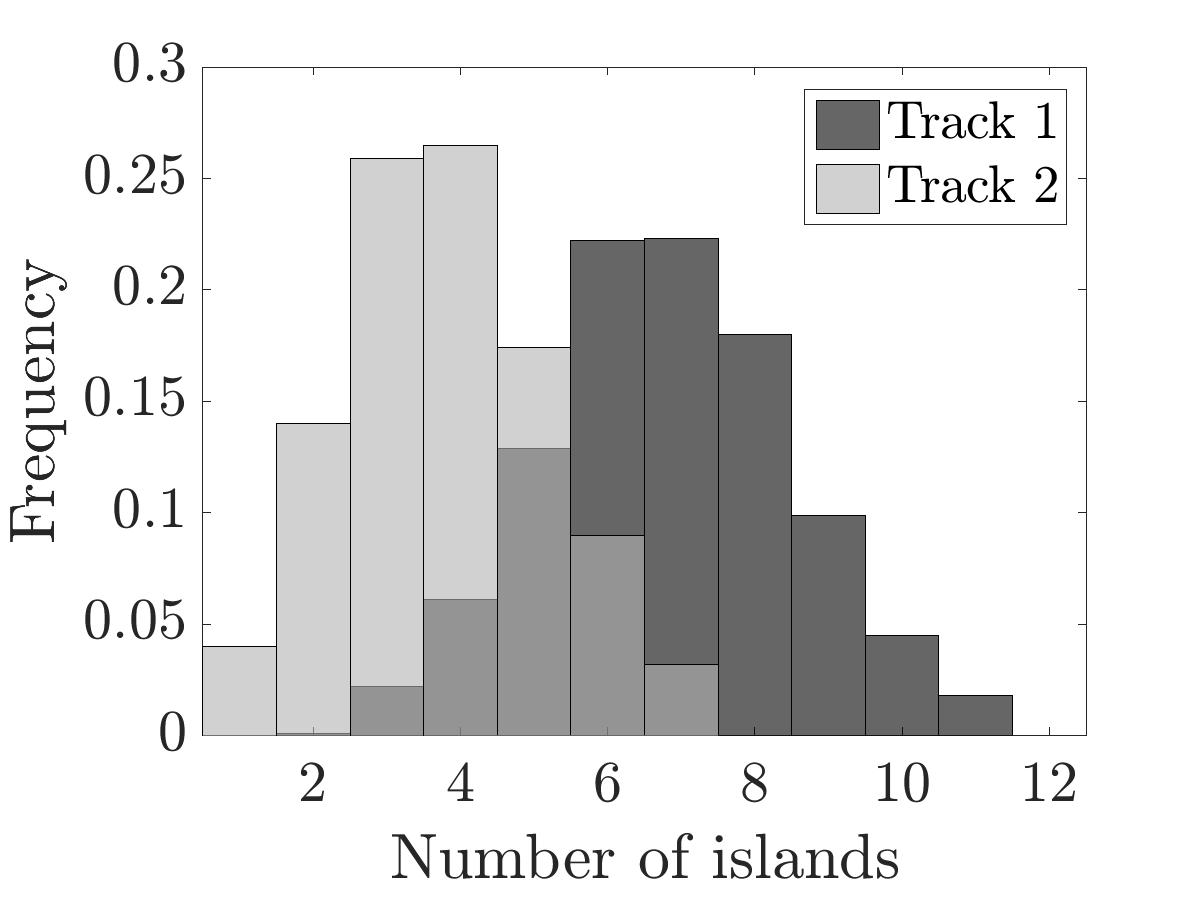}\label{fig:SimIslands}}
	\setcounter{subfigure}{0}	
	\caption{Empirical probability of number of line failures (\textit{left}) and number of islands formed (\textit{right}) in the 12-node network for the two tracks. A total of 1,000 failure scenarios are simulated to produce the histograms.}
	\label{fig:SimStats}
\end{figure}

We generate a total of $S=1000$ failure scenarios to examine distributions in frequency of failures and number/size of islands. The quadratic relationship between $\PPI{}{h,t}$ and $\vel{}{h,t}$ in \eqref{eq:PPI} results in a significant increase in probability of failures and decrease in island size if the test feeder is subject to higher storm velocities (see \cref{tabFailures}). \footnote{Category 2-5 storms have a larger radial region of high winds, and we can expect much higher failure probabilities in such cases. In contrast, under mild storms with the tropical storm rating, Poisson intensities are uniformly $\lambda_{norm}$, and the failure probability $\probFail{}{e}$ over $[\time_1,\time_2]$ is $<0.001$. } 
An average of 5.96 failures occur per scenario in Track 1, while only 2.88 failures occur per scenario in Track 2 (see \cref{fig:SimFailures}). The smaller median island size under Track 1 corresponds with a larger number of islands (see \cref{fig:SimIslands}). 

To implement the SAA solution approach, we utilize the mixed integer program (MIP) optimization model \eqref{eq:sooEmpirical} in JuliaPro. Solutions are obtained using the Gurobi solver.  

\subsection{Impact of resource constraints on DN recovery}
To evaluate the system performance under different resource constraints, we vary $\genBudget$ (number of available DERs) and $\crewBudget$ (maximum number of lines repaired in each period). System performance at a period $\period$ is defined as the average percentage difference between the cost of unmet demand and the total cost of complete load shedding, averaged over the sampled scenarios $\scenarioIdx$ and as a function of the optimal first-stage solution of each scenario $\hat\first_\scenarioIdx$, i.e. 
\begin{equation}
\begin{aligned}
\hspace{-0.2cm}\small\text{System performance} = 
\frac{1}{|\setScenariosSmall|}\sum_{\scenarioIdx \in \setScenariosSmall} 100\left(1-\frac{ \recourse(\hat\first_\scenarioIdx, \scenarioIdx)}{\sum_{i\in\N}\Cshed_i}\right). 
\end{aligned}
\end{equation}

\begin{figure}[htbp!]
	\subfloat{\includegraphics[width=0.5\textwidth, scale=0.12]{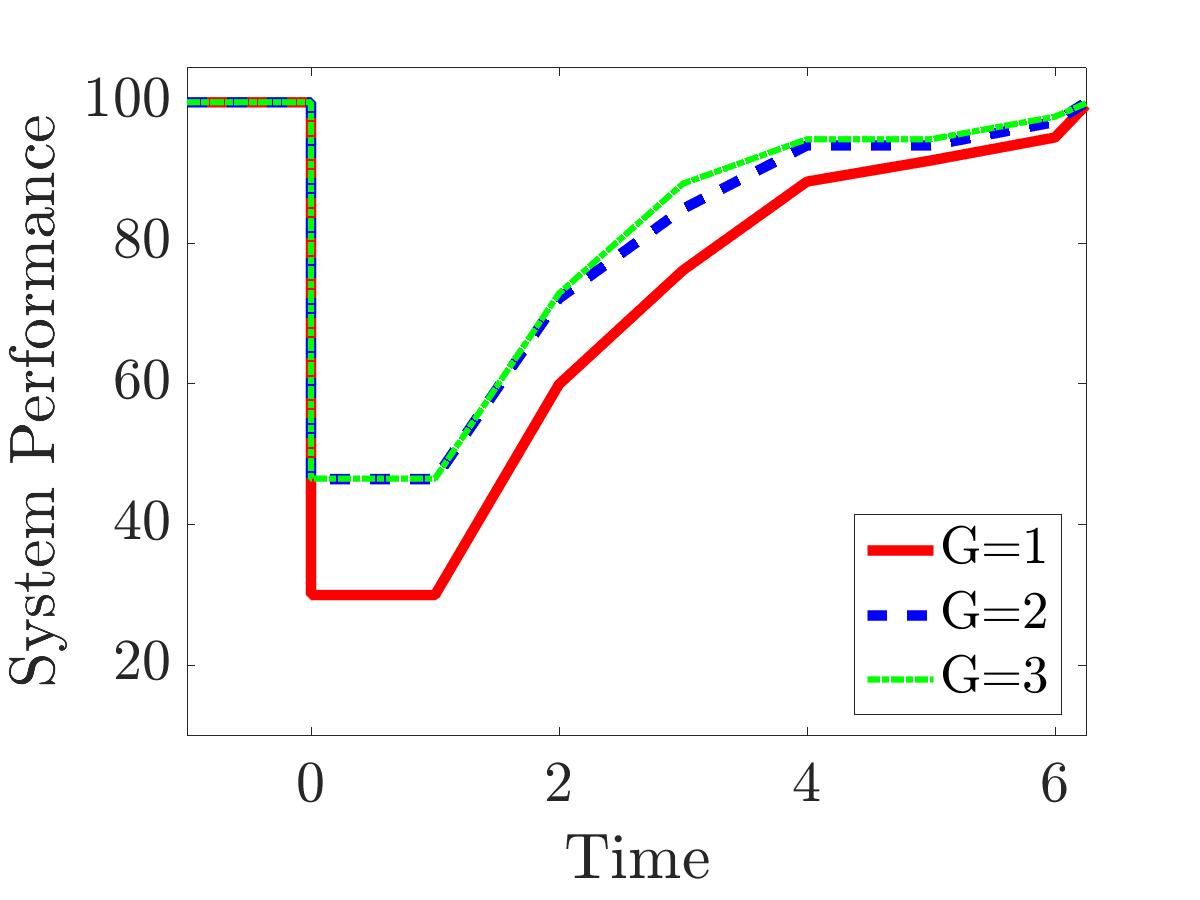}}
	\subfloat{\includegraphics[width=0.5\textwidth, scale=0.12]{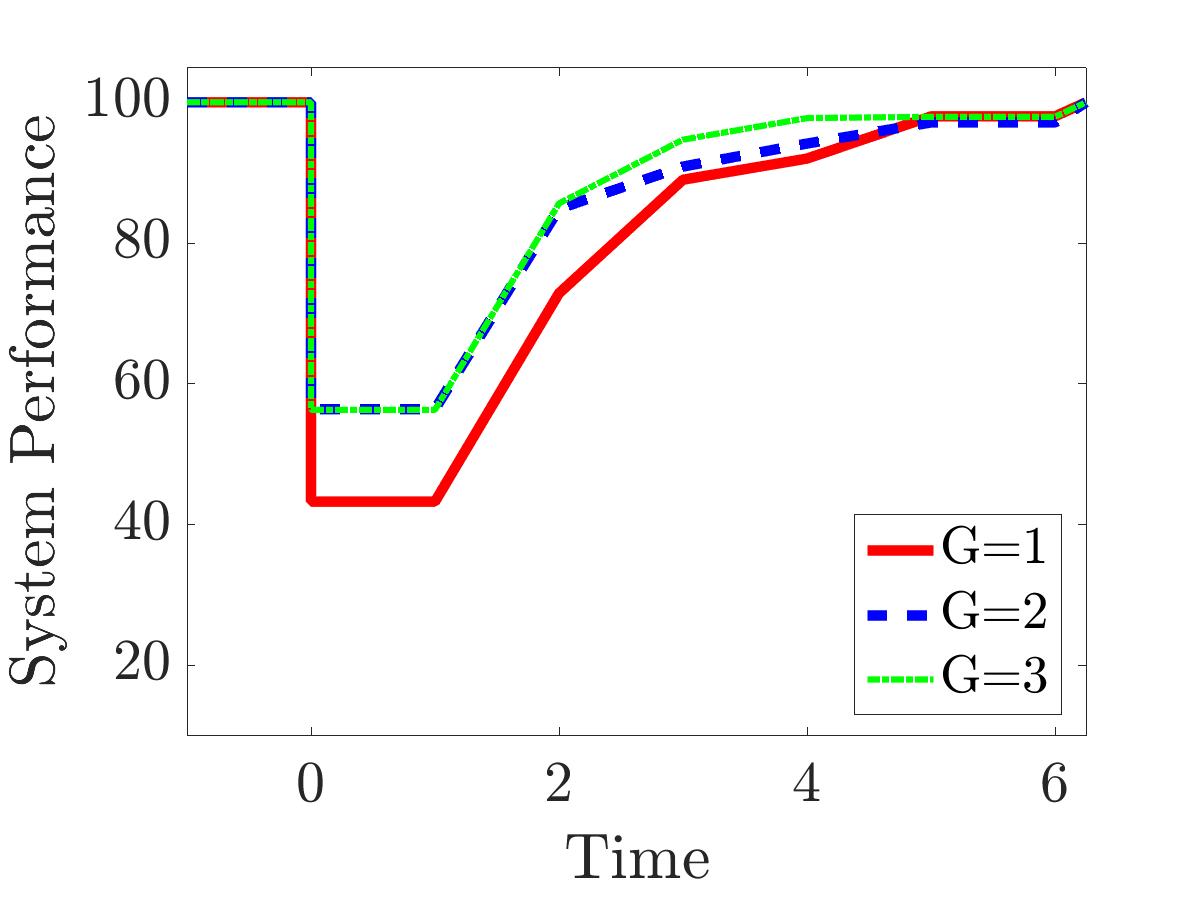}}
	\setcounter{subfigure}{0}%
	
	\subfloat[Track 1]{\includegraphics[width=0.5\textwidth, scale=0.12]{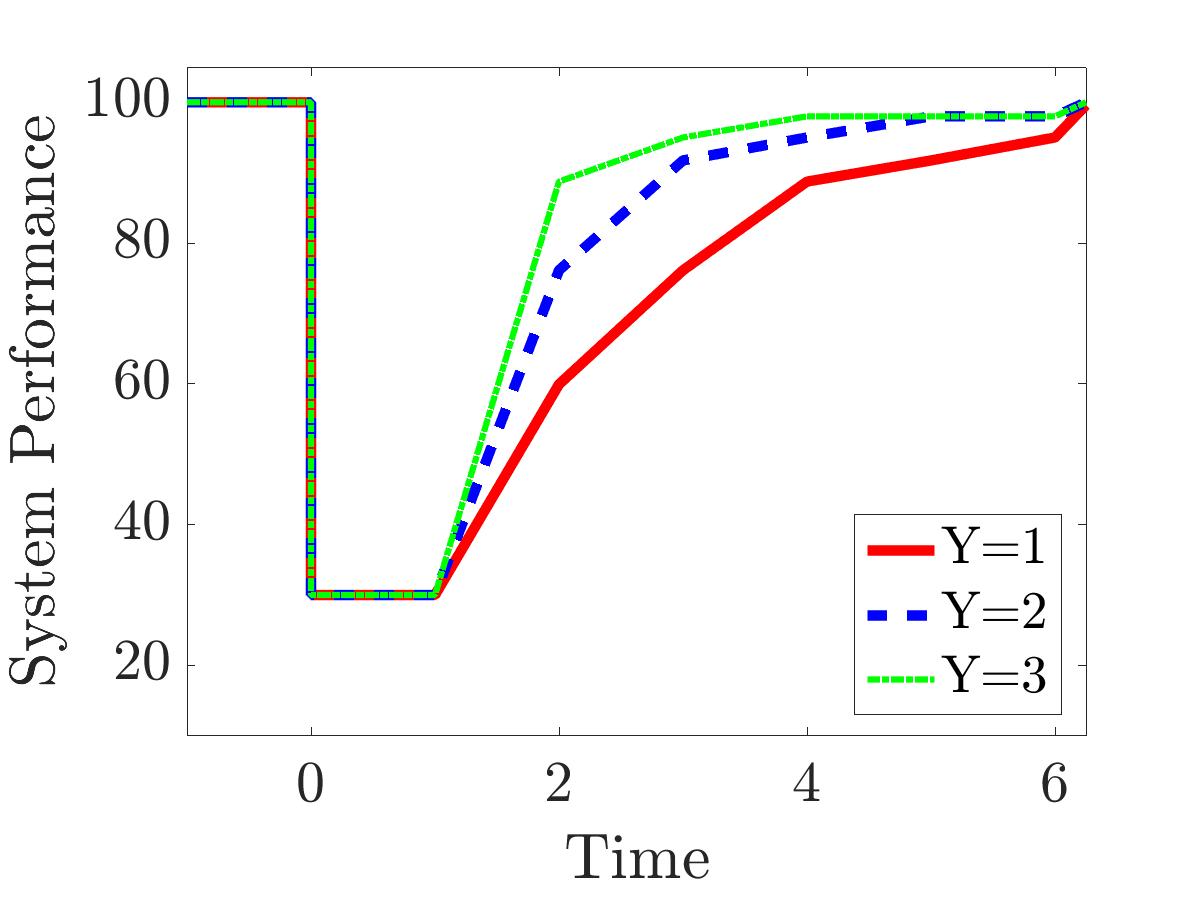}}
	\subfloat[Track 2]{\includegraphics[width=0.5\textwidth, scale=0.12]{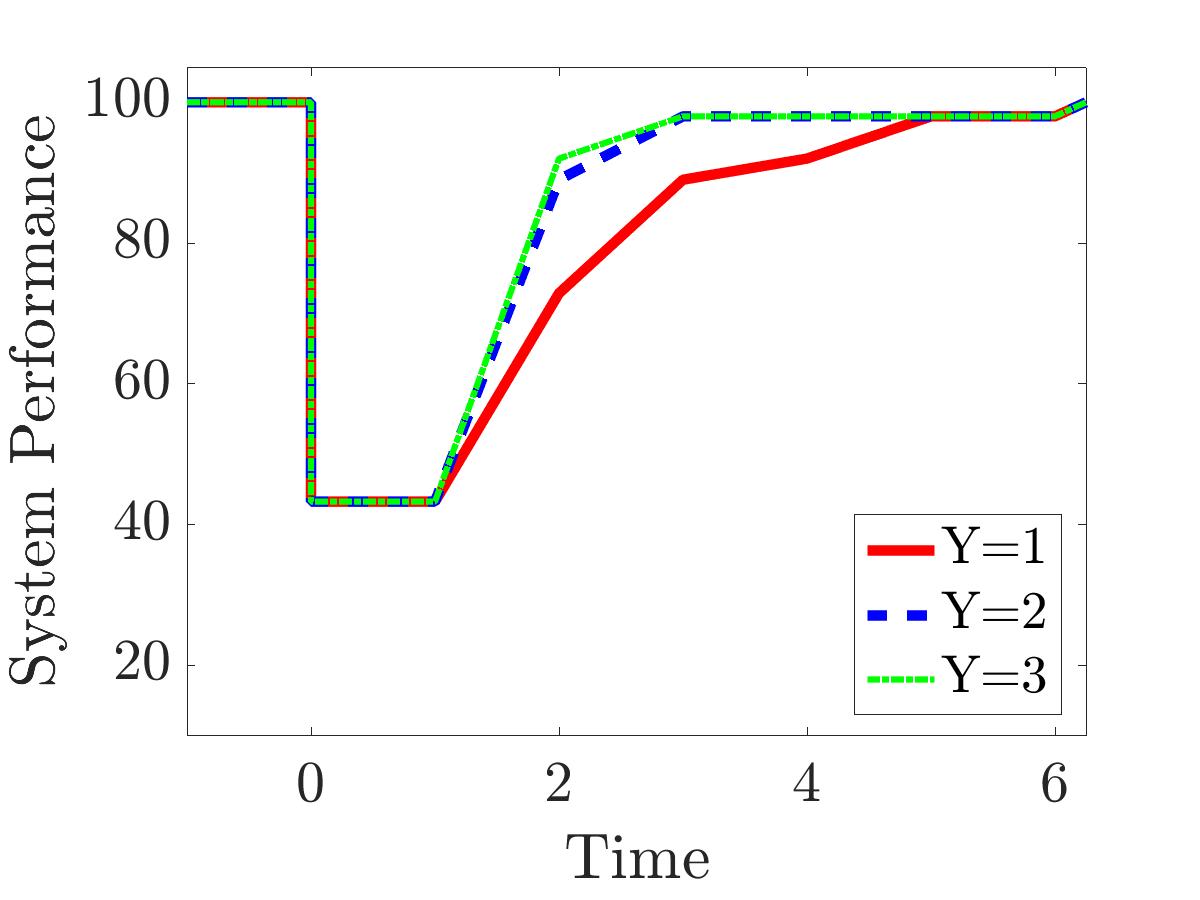}}
	\caption{Average system performance of the DN under the two track scenarios, varying $\protect\genBudget$ while setting $\protect\crewBudget=1$ (\textit{top row}) and varying $\protect\crewBudget$ while setting $\protect\genBudget=1$ (\textit{bottom row}).}
	\label{fig:resilienceCurves}
\end{figure}

The system performance as a function of $\period$ under different values of $\genBudget$ and $\crewBudget$ for two storm tracks is shown in \Cref{fig:resilienceCurves}. Following the storm ($\period=0$), the system performance is at a minimum, and improves with each subsequent set of line repairs. Once all the damaged lines are repaired, the system performance is almost (but, not fully) restored. The system performance is fully restored to 100\% following reconnection to the main grid. Since there are more failures on average in the DN under Track 1, the system performance at $\period=0$ is lower than for Track 2. 

If $\genBudget > 0$, even networks with high failure probabilities will be able to meet a portion of demand given a nonzero DER budget -- the network repair time will simply be longer. Increasing $\crewBudget$ noticeably decreases the average time required to repair the network (return to system performance that is close to 100\%) under both track scenarios. Increasing $\genBudget$ ensures that a larger portion of system demand is met while line repairs are not yet completed. 

\section{Concluding remarks}\label{sec:conludingRemarks}
We make the following contributions to address electricity network preparedness for storm-induced outages:
\begin{enumerate}
	\item Two-stage stochastic optimization formulation for DER placement and line repairs in DNs under uncertainty in component failure locations,
	\item Nonhomogeneous Poisson process (NHPP) model to predict spatially-varying likelihood of line failures, and
	\item Model of post-storm microgrid operation with DERs.
\end{enumerate}

Future work will focus on improving computational aspects of our solution approach. As stated in \cref{sec:lineFailureModel}, the sample $\setScenariosSmall$ used for SAA may not approximate the probability distribution $\probFails$ well. To obtain $\setScenariosSmall$ that is representative of $\probFails$, one can use a scenario reduction method such as the forward selection or backward reduction algorithm \cite{scenarioReduction}. The quality of solutions can then be evaluated by calculation of the optimality gap \cite{KleywegtShapiro}.

To decrease computation time, we will consider greedy heuristics to solve the multi-stage Stage II decision problem. Specifically, one can select a subnetwork with a small number of nodes based on their criticality. A node has high criticality if restoration of its load has high benefit on other intermediate nodes. The relative weights of the selected nodes can be determined based on their criticality. Running the SAA method on the (minimum spanning tree) subnetwork induced by this smaller set of nodes can lead to a computational speed-up that allows for testing the model on larger DN feeders.

\bibliography{sources/IEEEfull3.bib}
\end{document}

%% file: sources/preamble.tex
\usepackage{etex}
\reserveinserts{28}
\usepackage{makeidx}         
\usepackage{graphicx}        
\usepackage{multicol}        
\usepackage{multirow}
\usepackage[bottom]{footmisc}

\makeindex             

\usepackage{enumitem, fancyhdr, amsthm, amsmath, amsfonts,  amssymb, csquotes, mathtools, url, graphicx, balance, comment, mathrsfs, upgreek, wrapfig, cases, pgfplots,  fancybox,  multirow, hhline, mathtools, chngcntr, booktabs, algpseudocode, mathabx, floatrow, algorithm, algorithmicx, mathrsfs, etex, etoolbox, xfrac,  cite, xstring, xparse, ifmtarg, xifthen, bm, array, multicol, layouts, printlen, color, xcolor, soul, cancel, tabularx, ragged2e}
\usepackage[caption=false]{subfig}
\usepackage{array,verbatim,multirow,bbm,float}
\usepackage[]{todonotes}
\usepackage[normalem]{ulem}

\usepackage{tikz}
\usetikzlibrary{decorations.pathreplacing}

\usetikzlibrary{patterns,arrows,shapes,snakes,automata,backgrounds,petri,plotmarks}

\usepackage[colorlinks=true,
            raiselinks=true,
            linkcolor=MidnightBlue,
            citecolor=Mahogany,
            urlcolor=ForestGreen,
            linkcolor=blue,
            citecolor=red,
            urlcolor=green,
            pdftitle={Optimal Resource Allocation in Electricity Distribution Networks During Contingencies},
            pdfkeywords={DN Security},
            pdfsubject={Initial Manuscript},
            plainpages=false,
            bookmarks=false,
            draft
            ]{hyperref}

\newcounter{algsubstate}


\usepackage[nameinlink]{cleveref}

\newif\ifLongVersion
\LongVersionfalse

\newif\ifArxivVersion
\ArxivVersionfalse

\newif\ifNotArxivVersion

\ifArxivVersion

\NotArxivVersionfalse
\else

\usepackage[nodisplayskipstretch]{setspace}

\NotArxivVersiontrue
\fi

\crefname{section}{Sec.}{Sec.}
\Crefname{section}{Sec.}{§§}
\crefname{subsection}{\mytextsection}{\mytextsection\mytextsection}
\crefname{equation}{Eq.}{}
\Crefname{equation}{Eq.}{Eqs.~}

\newtheoremstyle{theoremdd}
{5pt}
{5pt}
{\itshape}
{0pt}
{\bfseries}
{.}
{ }
{\thmname{#1}\thmnumber{ #2}\thmnote{ (#3)}}

\theoremstyle{theoremdd}

\newtheorem*{formulation*}{}

\theoremstyle{remark}

\crefname{table}{Table}{Tables}
\crefname{figure}{Figure}{Figures}
\crefname{theorem}{Theorem}{Theorems}
\crefname{proposition}{Proposition}{Propositions}
\crefname{lemma}{Lemma}{Lemmas}
\crefname{algorithm}{Algorithm}{Algorithms}
\crefname{myclaim}{Claim}{Claims}

\makeatletter
\def\thm@space@setup{%
  \thm@preskip=0.2cm plus 0.2cm minus 0.2cm
  \thm@postskip=\thm@preskip 
}
\renewenvironment{proof}[1][\proofname]{\par
  \pushQED{\qed}%
  \normalfont
  \topsep0pt \partopsep5pt 
  \trivlist
  \item[\hskip\labelsep
        \itshape
    #1\@addpunct{.}]\ignorespaces
}{%
  \popQED\endtrivlist\@endpefalse
  \addvspace{6pt plus 6pt} 
}
\makeatother

\makeatletter

\newtheoremstyle{named}{}{}{\itshape}{}{\bfseries}{.}{.5em}{\thmnote{#3's }#1}
\theoremstyle{named}

\newtheoremstyle{mynamed}{}{}{\itshape}{}{\bfseries}{.}{.5em}{#1 \thmnote{A }}
\theoremstyle{named}

\newcommand{\abs}[1]{\left\lvert{#1}\right\rvert}

\allowdisplaybreaks

\counterwithout{equation}{section} 
\counterwithout{figure}{section} 

\algrenewcommand{\algorithmiccomment}[1]{\hskip3em$\slash\slash$ #1}

\def \setDER {\mathcal{D}}
\def \der {d}
\def \ref {\text{ref}}
\def \site {i}
\def \setSite {\mathcal{U}}
\def \setScenarios {\mathcal{S}}
\def \setScenariosSmall {\widehat{\setScenarios}}

\newcommand{\N}[1][]{\mathcal{N}_{#1}}
\newcommand{\E}[1][]{\mathcal{E}_{#1}}
\newcommand{\G}[1][]{\mathcal{G}_{#1}}

\newcommand{\fromNode}[1]{{#1}^{-}}
\newcommand{\toNode}[1]{{#1}^{+}}

\def \M {\mathcal{K}}
\def \bigM {\mathrm{L}}

\def \R {\mathbb{R}}

\def \x {{x}}

\def \j {\mathbf{j}}

\def \state {\eta}
\def \opr {o}
\def \con {c}
\def \attack {a}
\def \optimalAttack {oa}
\def \defend {d}
\def \hardMin {hmin}
\def \hardMax {hmax}

\def \NN {{\mathrm{N}}}

\def \C {{\text{W}}}

\def \Csite {\C^\text{\footnotesize SD}}
\def \Cshed {\C^\text{\footnotesize LS}}
\def \Cload {\C^\text{\footnotesize LC}}

\def \Xnpf {\mathcal{X}}

\def \first {a}
\def \First {\mathcal{A}}

\newcommand{\resistance}[1]{\text{r}_{#1}} 
\newcommand{\reactance}[1]{\text{x}_{#1}} 
\newcommand{\impedance}[1]{\text{z}_{#1}}

\def \ssum {\textstyle\sum}
\def \pprod {\textstyle\prod}

\usepackage{chngcntr}
\counterwithout{subsubsection}{subsection}

\makeatletter
\@addtoreset{subsubsection}{subsection}
\makeatother

\usepackage{tcolorbox}

\usepackage{mdframed}

\def \nperiod {\mathrm{K}}
\def \nscenario {S}
\def \period {k}
\def \scenario {S}
\def \time {t}
\def \scenarioIdx {s}
\def \node {i}
\def \ts {\period,\scenarioIdx}
\def \pre {pre}
\def \post {post}
\newcommand{\myc}[3]{
	\def \firstString {{#1}}	
	\IfEqCase{#2}{
		{}{\firstString_{#3}}
		{t}{\firstString_{#3}^{t}}
		{T}{\firstString_{#3}^{T}}
		{\site}{\firstString_{#3}^{\site}}
		{\ts}{\firstString_{#3}^{\ts}}
		{\scenarioIdx}{\firstString_{#3}^{\scenarioIdx}}
		{\ts-1}{\firstString_{#3}^{\ts-1}}
		{\period}{\firstString_{#3}^{\period}}
		{\period-1}{\firstString_{#3}^{\period-1}}
		{\period,\scenarioIdx}{\firstString_{#3}^{\period,\scenarioIdx}}
		{\nperiod,\scenarioIdx}{\firstString_{#3}^{\nperiod,\scenarioIdx}}
		{\period-1,\scenarioIdx}{\firstString_{#3}^{\period-1,\scenarioIdx}}
		{0}{\firstString_{#3}^{0}}
		{\opr}{\firstString_{#3}^{\opr}}
		{\con}{\firstString_{#3}^{\con}}
		{\star}{\firstString_{#3}^{\star}}
		{n}{\firstString_{#3}}
		{l}{\widehat{\firstString}_{#3}}
		{u}{\widecheck{\firstString}_{#3}}
		{max}{\mathbf{\overline{\firstString}}_{#3}}
		{min}{\mathbf{\underline{\firstString}}_{#3}}
		{\hardMax}{\bm{\overline{\overline{\firstString}}_{#3}}}
		{\hardMin}{\bm{\underline{\underline{\firstString}}_{#3}}}	
		{constant}{\mathbf{\firstString}_{#3}}
		{\pre}{\firstString_{#3}^{n}}
		{\post}{{\firstString}_{#3}^{c}}
		{act}{\firstString_{#3}^{act}}
		{set}{\firstString_{#3}^{set}}
		{ref}{\firstString_{#3}^{\text{ref}}}
		{r}{\firstString_{#3}^{r}}
		{t-1}{\firstString_{#3}^{t-1}}
		{nom}{\mathbf{\firstString_{#3}^{nom}}}
		{stab}{\bm{\firstString_{#3}^{stab}}}
		{devmax}{\firstString_{#3}^{dev,max}}
		{reg}{\bm{\firstString_{#3}^{reg}}}
		{ev}{\firstString_{#3}^{ev}}
		{nev}{\firstString_{#3}^{nev}}
		{maxev}{\bm{\overline{\firstString}_{#3}^{ev}}}
		{maxnev}{\bm{\overline{\firstString}_{#3}^{nev}}}
		{\state}{\firstString_{#3}^{\state}}
		{\attack}{\firstString_{#3}^{\attack}}
		{\optimalAttack}{\firstString_{#3}^{\attack\star}}
		{\defend}{\firstString_{#3}^{\defend}}
	}[\firstString_{#3}^{#2}]
	{
	}
}
\newcommand{\pfc}[2]{\myc{\eta}{#1}{#2}}	

\newcommand{\kcc}[2]{\myc{kc}{#1}{#2}}	
	
\newcommand{\klinec}[2]{\myc{kl}{#1}{#2}}

\newcommand{\nuc}[2]{\myc{\mathrm{v}}{#1}{#2}}

\newcommand{\Pc}[2]{\myc{P}{#1}{#2}}
\newcommand{\Qc}[2]{\myc{Q}{#1}{#2}}

\newcommand{\nucc}[2]{\myc{vc}{#1}{#2}}

\newcommand{\pcc}[2]{\myc{pc}{#1}{#2}}
\newcommand{\qcc}[2]{\myc{qc}{#1}{#2}}

\newcommand{\ptc}[2]{\myc{p}{#1}{#2}}
\newcommand{\qtc}[2]{\myc{q}{#1}{#2}}

\newcommand{\kqc}[1]{\myc{mq}{constant}{#1}}

\newcommand{\ysc}[2]{\myc{u}{#1}{#2}}
\newcommand{\ygc}[2]{\myc{w}{#1}{#2}}
\newcommand{\ylinec}[2]{\myc{yl}{#1}{#2}}

\newcommand{\pgc}[2]{\myc{pg}{#1}{#2}}
\newcommand{\qgc}[2]{\myc{qg}{#1}{#2}}

\newcommand{\xc}[2]{\myc{\x}{#1}{#2}}
\newcommand{\yc}[2]{\myc{y}{#1}{#2}}

\newcommand{\Xc}[2]{\myc{\Xnpf}{#1}{#2}}
\newcommand{\Yc}[2]{\myc{\mathcal{Y}}{#1}{#2}}
\newcommand{\lcc}[2]{\myc{\beta}{#1}{#2}}

\newcommand{\mycc}[4]{
	\def \firstString {{#1}}	
	\IfEqCase{#2}{
		{n}{\firstString_{#3}^{#4}}
		{max}{\overline{\firstString}_{#3}^{#4}}
		{min}{\underline{\firstString}_{#3}^{#4}}
	}
}

\newcommand{\soo}{\myc{g}{}{}}
\newcommand{\costSO}{\myc{\mathrm{W}}{}{}}
\newcommand{\costFirstStage}{\myc{\costSO_{\text{alloc}}}{}{}}
\newcommand{\costSecondStage}{\myc{\costSO_{\text{dem}}}{}{}}
\newcommand{\probFails}{\myc{\mathcal{P}}{}{}}
\newcommand{\recourse}{\myc{J}{}{}}
\newcommand{\uncertaintyR}{\myc{S}{}{}}
\newcommand{\uncertaintyT}[2]{\myc{\xi}{}{}}

\newcommand{\genBudget}{\myc{\mathrm{G}}{}{}}
\newcommand{\crewBudget}{\myc{\mathrm{Y}}{}{}}

\newcommand{\vel}[2]{\myc{v}{#1}{#2}}  
\newcommand{\tstorm}[1]{\myc{t}{#1}{}}  
\newcommand{\tstormS}{\myc{\time}{}{1}} 
\newcommand{\tstormF}{\myc{\time}{}{2}}  
\newcommand{\velCrit}{\myc{v_{crit}}{}{}}
\newcommand{\radius}[2]{\myc{r}{#1}{#2}}  
\newcommand{\Vm}{\myc{V}{}{\time}}  
\newcommand{\Rm}{\myc{R}{}{\time}}  
\newcommand{\B}{\myc{B}{}{\time}}  

\newcommand{\NHPPscale}{\myc{\alpha}{}{}}
\newcommand{\grid}{\myc{h}{}{}}
\newcommand{\gridSet}{\myc{\ch}{}{}}
\newcommand{\PPI}[2]{\myc{\lambda}{#1}{#2}}
\newcommand{\PPInorm}{\myc{\lambda_{norm}}{}{}}
\newcommand{\CDF}[2]{\myc{\Lambda}{#1}{#2}}
\newcommand{\CDFEdge}[2]{\myc{\nu}{#1}{#2}}
\newcommand{\lineLength}[2]{\myc{l}{#1}{#2}}
\newcommand{\probFail}[2]{\myc{F}{#1}{#2}}
\newcommand{\probExp}{\mathbf{Pr}}

%% file: sources/macros.tex
       \newcommand{\ch}{\mathcal{H}}                 

